# CHAPTER 9

# LEVERAGING SOCIAL MEDIA DATA FOR COVID-19 STUDIES

Nur Hafieza Ismail, Nur Shazwani Kamarudin, Nurol Husna Che Rose

## ABSTRACT

Nowadays, social media networks have become widely preferred sources of information. Especially during the time of the Coronavirus disease 2019 (COVID-19) pandemic, social media has been one of the most used platforms to get the latest news and information related to COVID-19. Social media is popular because it offers free access to its registered users and allows them to do posting, disseminate information and respond to others' postings. With almost 4.6 billion social media users worldwide, it is not surprising that the significant amount of information shared through these platforms could affect how people perceive and cope with the pandemic that we are facing right now. With decent use, social media can be a beneficial digital tool to spread reliable news and public awareness for patients, clinicians and society. Specifically, this chapter describes linguistic, visual and emotional indicators expressed in user disclosures. Thus, in this chapter, the related studies of social media platforms' usage during the COVID-19 pandemic are explored and discussed in detail. This chapter also categorises social media data used, and introduces different deployed machine learning, feature engineering, natural language processing, survey methods, and outlines directions for future research.

---

Nur Hafieza Ismail, Nur Shazwani Kamarudin,
Faculty of Computing, University Malaysia Pahang, 26600 Pekan, Pahang

Nurol Husna Che Rose
Faculty of Electronic Engineering Technology, University Malaysia Perlis, 02600 Arau, Perlis

Nur Hafieza Ismail (Corresponding Author)
e-mail: hafieza@ump.edu.my

**keywords** COVID-19, social media, linguistic data, visual data, machine learning

## INTRODUCTION

The COVID-19 pandemic has taken a number of lives around the world, where it first arose in Wuhan, China, at the end of the year 2019 [1]. Rapidly, the virus has spread all over the countries and forced the World Health Organization (WHO) to declare COVID-19 as a pandemic in March 2020. Many countries are implementing measures to help reduce the spreading of COVID-19 by enforcing the home quarantine, social distancing, and closure of many non-essential sectors from operating. Mass media, including social media platforms, played a major role in providing the latest information regarding COVID-19. In 2020, most people have very little knowledge about this virus. The widespread fake news, misinformation and rumours about COVID-19 across social media platforms cause panic and fear among readers. This messy situation could contribute to panic decisions, especially decisions taken for their health in preventing and curing the COVID-19 infection.

With social media, people can freely communicate and share their thoughts, ask personal questions, and seek peer support, especially for those with conditions that are highly stigmatised, without revealing personal identity. The fast spread of information related to COVID-19 on social media platforms, including Facebook, Twitter, Reddit, and Youtube, has been a growing concern of the governments and public health authorities. The misinformation on medical and unobjective content related to COVID-19 has been disseminated on social media at a brisk pace. It is a very urgent task to obstruct the fake news and rumours related to COVID-19 from spreading across social media platforms. If not, it could cause extreme fear, anxiety and worry among people, and without prevention action taken to stop this, these negative emotions growing in people can lead to mental health issues later. Thus, we have to accept that social media platforms are one of the most preferred mediums of information nowadays and we have to fully utilise this platform to convey reliable information and knowledge to the public. Some governments such as Malaysia have identified this problem and imposed new laws regarding fake news, where those who create, publish or share fake information on COVID-19 on any media platforms could face a fine not exceeding RM100,000, imprisonment not exceeding three years, or both [2].

Many scholars have studied and investigated COVID-19 information on social media platforms in recent years. The COVID-19 pandemic is addressed as the largest scientific experiment in human history in many academic disciplines, including medicine, psychology, economics, social, politics, and computer science. In this chapter, the related studies of social media as a source of information during the current global pandemic are discussed and explored. From the finding of recent work, the failure in promoting accurate practices to reduce the fake news spread has caused a man of a family with three children to commit suicide after being diagnosed with COVID-19. This tragic situation will not happen if people read and digest valid information from authentic sources [3]. Additionally, it is important to identify the possible reasons that lead people to keep sharing fake news over social media platforms. Thus, the main goal of this chapter is to pave the way for an open and ethical discussion about utilising social media data on studying COVID-19 and show how this new source of data can be tapped to improve medical practice, provide timely support, and influence government or policymakers.

## COVID-19

COVID-19 has produced a world health crisis and has had big changes in how we live our life. It is also known as a severe acute respiratory syndrome coronavirus 2 (SARS-CoV-2) that causes severe pneumonia with fatality risk [4]. This infectious disease has a high risk of transmission that causes high fatality risk in people. From the coronavirus resource centre website managed by Johns Hopkins University, until the end of February 2022, the record shows 430 million total cases and 5.92 million deaths due to COVID-19. This situation has caused economic repercussions, an increase in poverty and loss of human lives all across the world.

It became contagious, spread uncontrollably around the world and caused a killer epidemic at the time. The clinical early signs of COVID-19 that are frequently reported are fever, fatigue, and infections of the upper respiratory system. The affected person is required to be in an incubation period of an average of five days to prevent the spreading in the community. It may cause symptoms of severe pneumonia, leading to death [5]. The symptoms usually appear in less than a week and the infection can progress to a severe condition, which is pneumonia that had happened to 75% of the affected patients. Pneumonia symptoms usually appear during the second and third week of infection [6]. This has caused a physical threat to humans, and most governments around the world have enforced social and physical distancing. Thus, due to this new ruling, more people are seeking related information on social media and mass media channels.

## SOCIAL MEDIA ANALYSIS

Social media has become a good source for data collection. It can contribute to different kinds of fields that can give great benefits to people, regardless of whether they are using it or not. There are different types of data that can be used from social media, such as text, image, video and audio. The amount of data on social media data increases rapidly. For example, on Twitter, 350,000 tweets are generated per minute and 500 million tweets are generated per day. The majority of studies in social media analysis are not able to specify how spending time on social media may affect the mental health condition of the user. A major factor that might affect social media users is the way they use social media, because it can be very beneficial or toxic at the same time. For instance, active use of social media with two-way communication can be very beneficial to the user but it can also be destructive or toxic to the user.

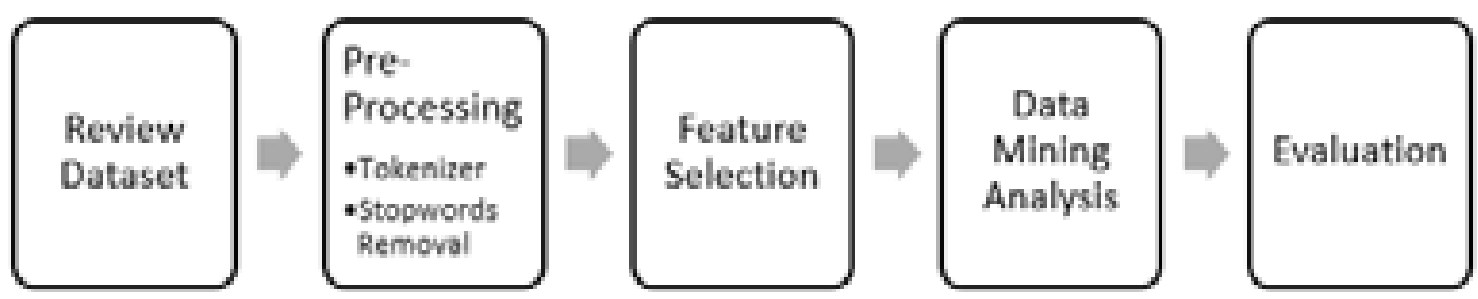

**Figure 1:** Steps of Social Media Analysis [7].

Figure 1 shows the common steps in social media analysis. It starts with a dataset review, in which the researchers need to choose the right dataset for their experiment. The second step is data pre-processing, which means preparing the data for the experiment, such as removing stop words or word/sentence tokenising. The next step is to select meaningful features from social media data such as an image or textual features. After selecting the right features, it is data mining analysis that includes deploying various techniques to develop the desired model. The final step is an evaluation, employing different metrics such as accuracy, recall, precision and F1 scores, for example.

## STUDYING COVID-19 IN SOCIAL MEDIA

As the social media data has recently emerged as the main medium to spread information among online communities, there are also various approaches used by researchers to study related problems. In this section, we elaborate on the types of techniques or tools used in their research. Next, we introduce a learning machine that is used for COVID-19 analysis in social media.

### Linguistic-based Data

Over the past few years, research in crisis informatics has utilised language as a medium to understand how major crisis events unfold in affected populations, and how they are covered in traditional media, as well as online media such as blogs and social media sites [8]. Interesting work by [9] introduced a specialised stress tweets classifier, which narrows down the theoretical algorithms to practical usage in the public health area and demonstrates more effective than traditional sentiment index classifiers that focus on the pattern of depressive symptoms caused by COVID-19. The Basilisk bootstrapping algorithm to find semantic lexicons could be used to divide the tweets into two categories: stressed and non-stressed. The author utilised Scikit-Learn TfidfVectorizer to transform preprocessed tweets into a sparse matrix. Then, the sparse matrix and lexicon are used by the anchored CorEx model to perform anchored topic modelling. Results show that the PHQ stress level in our results matches well with the number of increased cases illustrated in the Johns Hopkins Coronavirus Resource Centres' statistical analysis results. This means that the number of new cases reduced due to the social distancing practice, and at the same time, the level of people's major concerns in many geographic regions reduced as well.

Another work by [10] discussed social media in the times of COVID-19. This article basically explains how social networks have become central to the rapid dissemination of scientific information and administrative pandemic monitoring and control. The author also agrees that social media networks have the potential to lead the way forward by delivering valid scientific observations in an organised manner to the appropriate audience. Likewise, [11] discussed how the COVID-19 pandemic brought unforeseen challenges that could forever change the way societies prioritise and deal with public health issues. The approaches to contain the spread of the virus have entailed governments issuing recommendations on social distancing, lockdowns to restrict movements and suspension of services. Their work focus on online presence by children may have resulted in increased exposure to abusive content and cyberbullying. Conversations on Twitter were reviewed to measure increases in abusive or hateful content, and cyberbullying, while testimonials from Reddit forums were examined to monitor changes in references to family violence before and after the start of the stay-at-home restrictions. Their experiments found that violence-related subreddits were among the topics with the highest growth after the COVID-19 outbreak. The analysis of Twitter data shows a significant increase in abusive content generated during the stay-at-home restrictions.

The authors from [12] touch on the social media insights into US mental health during the COVID-19 pandemic by analysing Twitter data. They analysed 86,581,237 public domain English language US tweets collected from an open-access public repository in three main steps. First, they characterised the evolution of hashtags over time using latent Dirichlet allocation (LDA) topic modelling. Second, they increased the granularity of this analysis by downloading the Twitter timelines of a large cohort of individuals (n=354,738) in 20 major US cities to assess changes in social media use. Finally, using this timeline data, they examined collective shifts in public mood in relation to evolving pandemic news cycles by analysing the average daily sentiment of all timeline tweets with the Valence Aware Dictionary and Sentiment Reasoner (VADER) tool. LDA topics generated in the early months of the data set correspond to major COVID-19–specific events. However, as state and municipal governments began issuing stay-at-home orders, latent themes shifted toward US-related lifestyle changes rather than global pandemic-related events. Social media volume also increased significantly, peaking during stay-at-home mandates. Finally, VADER sentiment analysis scores of user timelines were initially high and stable but decreased significantly, and continuously, by late March.

[13] proposed a novel framework to analyse the topic and sentiment dynamics due to COVID-19 from the massive social media posts. In the proposed framework, they employ the Dynamic Topic Model (DTM) to generate accurate daily topics. To determine the sentiment polarity of each topic and tweet, they utilise a sentiment lexicon tool: VADER to infer the sentiment polarity. 13,746,822 tweets from 1 to 14 April 2020 related to COVID-19 from Twitter across the world were collected to test the effectiveness of the proposed framework. The experimental results show that the proposed framework can generate insightful findings such as the overall sentiment dynamics among people, topic evolutionary patterns and the sentiment dynamics of different topics. The results of the analysis showed that people are concerned about the latest confirmed coronavirus cases, measures to prevent infection, the attitudes and specific measures of governments towards the pandemic. The overall sentiment

polarity was positive, but topic sentiment polarity varies from topic to topic.

**Visual-based Data**

Imaging can contribute to so many fields, including COVID-19-related research. [14] worked on proposing a detection model using deep learning models. Images that were collected from Github were processed and then fed into the deep neural network. Given 260 images that were used in their study, the proposed model showed an impressive performance compared to human detection results. Another work by [15] studied the use and fit of face masks and social distancing in the United States and events of large physical gatherings through public social media images from 6 cities and Black Life Matter (BLM) protests. They collected and analysed 2.04 million public social media images from New York City, Dallas, Seattle, New Orleans, Boston and Minneapolis between 1 February 2020, and 31 May 2020.

[16] proposed a novel framework to collect, analyse, and visualise Twitter posts to specifically monitor the virus spread in Italy. The authors present and evaluate a deep learning localisation technique that geotags posts based on the locations mentioned in their text, a face detection algorithm to estimate the number of people appearing in posted images and a community detection approach to identify communities of Twitter users. From that data, the author proposes a further analysis of the collected posts to predict their reliability, and to detect trending topics and events. As for the face detection task, the Tinyfaces methodology was deployed and evaluated for the task of face counting. The results indicate better performance on small-scale faces in highly crowded scenes. Finally, the comparison of alternative community detection techniques by means of modularity and execution time showed the superiority of the Louvain algorithm that was adopted by the author.

## Combined Data of Linguistic and Visual-based

Apart from using only visual data or only linguistic data, researchers also combine these two kinds of data to study social media influences on mental health. Sociologists also claim that it is not possible to communicate by using only words; people also use pictures to communicate with each other [17]. Another work done by [18] that studied the effect of the social media environment on the perception of polarising topics is also being addressed also in the case of COVID-19. The authors provide an in-depth analysis of the social dynamics in a time window, where narratives and moods in social media related to the COVID-19 have emerged and spread. Data that are available on the WHO website were used for this specific work, which can be downloaded from https://covid19.who.int/data. This work analyses mainstream platforms such as Twitter, Instagram and YouTube, as well as less regulated social media platforms such as Gab and Reddit. They perform a comparative analysis of information spreading dynamics around the same argument in different environments having different interaction settings and audiences.

[19] identified five overarching public health themes concerning the role of online social media platforms and COVID-19. These themes focused on surveying public attitudes, identifying infodemics, assessing mental health, detecting or predicting COVID-19 cases, analysing government responses to the pandemic, and evaluating the quality of health information in prevention education videos. This work also emphasises the paucity of studies on the application of machine learning on data from COVID-19-related social media and a scarcity of studies documenting real-time surveillance that was developed with data from social media on COVID-19. Government responses that were distributed via social media have been increasingly crucial in combating infodemics, and promoting accurate and reliable information for the public. It also remained unknown whether government posts would reach greater numbers of social media users or would have greater effects on them than infodemics.

YouTube has served as one of the major platforms to spread information concerning the control of COVID-19.

## MACHINE LEARNING

The role of machine learning in the data science field has been proven useful through the use of statistical methods. Algorithms are trained to make classifications or predictions, and uncover key insights within data mining projects. Machine learning, deep learning and neural networks are all sub-fields of artificial intelligence. These machine learning techniques can be used specifically on the topic that we are discussing in this book chapter. Given the big social media data available out there, a researcher may be able to have a nugget discovery that can be useful for social and online communities.

[13] proposed a novel framework to analyse the topic and sentiment dynamics due to COVID-19 from the massive social media posts. Based on a collection of 13 million tweets related to COVID-19 over two weeks, they found that the positive sentiment shows a higher ratio than the negative sentiment during the study period. By utilising the Latent Dirichlet Allocation model for topic extraction and VADER sentiment analysis, they were able to extract the most interesting finding from the set of the tweet. When zooming into the topic-level analysis, they figured that different aspects of COVID-19 have been constantly discussed and show comparable sentiment polarities. On the other hand, another interesting research by [9] was done on misinformation detection using social media data. They even released a dataset for other researchers to use, known as COVIDLIES, that is available for the public to use and can be accessed here (https://ucinlp.github.io/covid19/). This dataset contains 6761 expert-annotated tweets to evaluate the performance of misinformation detection systems on 86 different pieces of COVID-19-related misinformation. The stance detection sub-task is a standard classification problem with three classes (Agree, Disagree and No Stance). Then, the author evaluates by measuring the precision, recall and F1 score of the predicted classes.

BERTSCORE (DA) + SBERT (DA) (on MultiNLI) achieves the highest F1 (41.2%) for the Agree class, while also obtaining the highest macro averaged Precision (55.9%) and F1 (50.2%). The combined BERTSCORE (DA) + NLI approach, in general, improves F1 across all classes for all models.

Another work by [20] aimed to design text mining to detect anxiety during a pandemic by applying machine learning technology. Two methods of machine learning were used in this work, which are Random Forest and Extreme Gradient Boosting (XGBoost). This work uses a sample of data from YouTube comments with a total of 4,862, consisting of 3,211 negative data and 1,651 positive data. This research concept applies systems development technology using prototyping techniques to model the detection of sentiment and emotional opinions. The data collected were preprocessed for normalisation and detection. The machine learning process is used to detect the sentiment (anxiety) from comments. Based on the results and discussions that have been done, the recommended machine learning method is the Random Forest method, which has an accuracy of more than 83% compared to XGBoost, which only has 73% accuracy.

## DISCUSSION

As the global COVID-19 pandemic continues to challenge societies across the globe and as access to accurate information about the virus itself. The COVID-19 pandemic brought unforeseen challenges that could forever change the way societies prioritise and deal with public health issues. Studying what people are posting on social media can shed a light on the policy maker and government in taking the right actions to protect our society. This chapter presents an overview of COVID-19 studies related to work that uses social media data and machine learning in their work. This chapter has discussed several works that use text data, image data or a combination of both in studying social media. It shows us how we can leverage the rich amount of data available out there and make it a good use in helping people in need. We discuss three key points focusing on COVID-19 studies using social media data, namely data, approaches, and findings. We found that the majority of work that is studying about COVID-19 using social media data was related to mental well-being. Given text data from Reddit, Twitter, Facebook and Youtube, a researcher in this field was able to do early detection on people that were suffering during this pandemic season. The use of machine learning was found to be useful for this kind of study. There is a lot of work proving that leveraging machine learning methods have improved the learning process, especially while working with a big set of data from social media.